\documentclass[10pt]{iopart} 
\usepackage{iopams}
\usepackage{setstack}
\usepackage{amsthm}
\usepackage{bbm}
\usepackage{psfrag}
\usepackage[dvips]{graphicx}
 
\newcommand\NN{{\mathbbm{N}}}
\newcommand\RR{{\mathbbm{R}}}

\newcommand{\Heaviside}{\Theta}
\newcommand{\Dirac}{\delta}

\newcommand{\Ima}{\mbox{\rm Im}}

\newcommand{\Ci}{\mbox{\rm Ci}}
\newcommand{\Si}{\mbox{\rm Si}}
\newcommand{\mod}{\mbox{\rm mod}}

\newtheorem{proposition}{Proposition}
\newtheorem{theorem}{Theorem}

\begin{document} 

\noindent FAU--TP3--07/4

\title[Nonanalyticities of the entropy induced by saddle points]{Nonanalyticities of the entropy induced by saddle points of the potential energy landscape} 

\author{Michael Kastner$^1$, Oliver Schnetz$^2$ and Steffen Schreiber$^1$} 
\address{$^1$ Physikalisches Institut, Universit\"at Bayreuth, 95440 Bayreuth, Germany}
\address{$^2$ Institut f\"ur Theoretische Physik III, Friedrich-Alexander-Universit\"at Erlangen-N\"urnberg, Staudtstra{\ss}e 7, 91058 Erlangen, Germany}
\ead{Michael.Kastner@uni-bayreuth.de} 

\date{\today}
 
\begin{abstract}
The relation between saddle points of the potential of a classical many-particle system and the analyticity properties of its Boltzmann entropy is studied. For finite systems, each saddle point is found to cause a nonanalyticity in the Boltzmann entropy, and the functional form of this nonanalytic term is derived for the generic case of potentials having the Morse property. With increasing system size the order of the nonanalytic term grows unboundedly, leading to an increasing differentiability of the entropy. Nonetheless, a distribution of an unboundedly growing number of saddle points may cause a phase transition in the thermodynamic limit. Analyzing the contribution of the saddle points to the density of states in the thermodynamic limit, conditions on the distribution of saddle points and their curvatures are derived which are necessary for a phase transition to occur. With these results, the puzzling absence of topological signatures in the spherical model is elucidated. As further applications, the phase transitions of the mean-field $XY$ model and the mean-field $k$-trigonometric model are shown to be induced by saddle points of vanishing curvature. 
\end{abstract}
%\pacs{05.70.Fh, 05.20.-y, 75.10.Hk} 


\section{Introduction}

Phase transitions are abrupt changes of the macroscopic properties of many-particle systems under variation of a control parameter. Typical examples are the sudden disappearance of the electric resistance when cooling a superconducting material below its transition temperature, or the evaporation of a liquid at temperatures above its boiling point. An approach commonly used for the theoretical study of phase transitions is the investigation of the analyticity properties of thermodynamic functions like the canonical free energy of enthalpy. It is long known that non\-an\-a\-lyt\-ic behaviour in a canonical thermodynamic function can occur only in the thermodynamic limit in which the number of degrees of freedom $N$ of the system goes to infinity \cite{Griffiths}.

Many researchers took it for granted that the same were true also for mi\-cro\-ca\-non\-i\-cal thermodynamic functions. Recently, however, it was observed that the mi\-cro\-ca\-non\-i\-cal entropy, or Boltzmann entropy, of a finite system is not necessarily real-analytic, i.\,e., not necessarily infinitely many times differentiable. Nonanalytic entropy functions of finite systems have been reported for certain classical models \cite{KaSchne:06,DunHil:06,HilDun:06,CaKa:06} as well as for quantum systems \cite{BroHoHu:07}, where the latter result relies on a suitable but rather unconventional definition of the density of states.

In light of their conceptual similarity to the definition of phase transitions, it is tempting to regard finite-system nonanalyticities of the entropy as phase transition-like phenomena. This point of view is advocated in \cite{DunHil:06}, and the authors of that reference argue that such nonanalyticities should also be measurable experimentally, at least in very small systems. However, such an interpretation is complicated by the fact that, as discussed in \cite{CaKa:06}, for typical models the number of nonanalytic points of the entropy increases unboundedly with the number of degrees of freedom $N$. 

Due to their typically large number, one might assume that nonanalyticities of the finite-system entropy were {\em unrelated}\/ to the occurrence of a phase transition in the thermodynamic limit. From a theorem by Franzosi and Pettini \cite{FraPe:04}, however, it can be deduced that---at least for a certain class of short-range systems---a relation between finite-system and infinite-system nonanalyticities exists (and we will come back to that theorem later). The purpose of the present article is to further clarify and quantify the relation between saddle points of the potential energy and nonanalyticities of the entropy for the cases of finite as well as infinite systems. For infinite systems, at least in the common situation of ensemble equivalence, a nonanalyticity of the entropy will correspond to a nonanalyticity in the canonical free energy, and hence to a phase transition according to the standard definition. Note that in the literature other approaches relating saddle points of the potential to phase transitions have been proposed. Examples include the work of Rutkevich \cite{Rutkevich:92}, Wales \cite{Wales}, and Pettini \cite{Pettini}, and we will elaborate on the relation to the latter approach in Sec.~\ref{sec:tdl}.

Our approach is the following: After fixing notations in Sec.~\ref{sec:preliminaries}, we discuss the analyticity properties of the Boltzmann entropy for a generic class of classical many-particle systems in Sec.~\ref{sec:finite}. For the case of finite systems, our main result is that every saddle point (including maxima or minima) of the high-dimensional potential energy landscape corresponds to a nonanalyticity of the Boltzmann entropy, and the functional form of this nonanalytic term is derived explicitly. (A weaker version of this result has been announced without proof in a recent Letter \cite{KaSchneSchrei:07}). This result gives a model-independent and quantitative account of nonanalyticities in the finite-system entropy as observed for the special cases in \cite{KaSchne:06,DunHil:06,CaKa:06}. Then, the density of states is split into two terms: The first is a sum of the nonanalytic contributions stemming from the saddle points, whereas the second contains the analytic rest. Performing the thermodynamic limit of the thus obtained expressions, we obtain in Sec.~\ref{sec:tdl} a condition on the distribution of saddle points and their curvatures which necessarily has to be fulfilled in order cause a phase transition. (This result has been outlined in the recent Letter \cite{KaSchne:08}.) We apply our findings to the spherical model of a ferromagnet in Sec.~\ref{sec:spherical}, explaining the puzzling behaviour that, for this model, the topological signatures observed do not coincide with the phase transition energy \cite{RiRiSta:05,RiRiSta:06}. As further examples, the mean-field $XY$ model (Sec.~\ref{sec:XY}) and the mean-field $k$-trigonometric model (Sec.~\ref{sec:k-trig}) are discussed, illustrating nicely the relation of curvature properties of saddle points of the potential and the occurrence of phase transitions in the thermodynamic limit. We summarize our findings in Sec.~\ref{sec:summary}.

\section{Preliminaries}
\label{sec:preliminaries}

We consider classical systems of $N$ degrees of freedom, characterized by Hamiltonian functions of standard form,
\begin{equation}\label{eq:Hamiltonian}
H(p,q)=\frac{1}{2}\sum_{i=1}^N p_i^2 + V(q), 
\end{equation}
where $p=(p_1,\dots,p_N)$ is the vector of momenta and $q=(q_1,\dots,q_N)$ the vector of position coordinates. The restriction to a quadratic form in the momenta is not essential, but simplifies the presentation. The potential $V$ is an analytic mapping from the configuration space $\Gamma_N\subseteq\RR^N$ onto the reals. Whenever $\Gamma_N$ is noncompact, we assume $V$ to be {\em confining}, i.\,e.,
\begin{equation}
\lim_{\lambda\to\infty}V(\lambda q)=\infty \qquad\forall 0\neq q\in\Gamma_N.
\end{equation}

In general $V$ will have a number of {\em critical points}\/ (or {\em saddle points}) $q_\mathrm{c}$, defined as points from $\Gamma_N$ with vanishing differential, $\rmd V(q_\mathrm{c})=0$. Note that throughout this article the expression ``saddle points'' is used synonymously to ``critical points'' and includes also minima and maxima. We will assume in the following that all critical points of $V$ are non-degenerate, i.\,e., that the determinant of the Hessian ${\mathfrak H}_V$ fulfills
\begin{equation}\label{eq:Morse}
\det \left[{\mathfrak H}_{V}(q_\mathrm{c})\right]\neq0\qquad\mbox{$\forall q_\mathrm{c}$ of $V$}.
\end{equation}
In this case, $V$ is called a {\em Morse function}. Conceptually, this is an insignificant restriction, since Morse functions on some manifold $M$ form an open dense subset in the space of smooth functions on $M$ \cite{Demazure,Hirsch}. Therefore, if the potential $V$ of the Hamiltonian system is not a Morse function, we can deform it into a Morse function $\bar{V}$ by adding an arbitrarily small perturbation, e.g.
\begin{equation}\label{eq:perturbation}
\bar{V}(q)=V(q)+\sum_{i=1}^N h_i q_i
\end{equation}
with small $h_i\in\RR$ ($i=1,\dots,N$). For practical purposes, adding a perturbation---and thereby destroying a symmetry present in $V$---may, however, render the computation of critical points and indices more complicated or even impossible.

In case $V$ is not a Morse function due to the presence of a symmetry, an alternative strategy is possible: zero-eigenvalues of the Hessian caused by, say, translation invariance can be removed by fixing one or several position coordinates, thereby destroying the symmetry. Thermodynamic functions of such a system with a reduced number of degrees of freedom will differ from the original ones only by a physically irrelevant additive constant.

\section{Finite system nonanalyticities}
\label{sec:finite}

When computing thermodynamic properties of a Hamiltonian system, a quadratic form in the momenta as in the Hamiltonian \eref{eq:Hamiltonian} merely leads to a shift in the free energy. Hence, for the discussion of phase transitions or nonanalyticities of thermodynamic function we may disregard the kinetic term and concentrate on the configurational part (see \cite{CaKa:07} for a discussion of the pitfalls of this reasoning).

Our aim is to investigate the analyticity properties of the configurational density of states as a function of the potential energy per degree of freedom $v=V/N$,
\begin{equation}
\Omega_N(v)=\int_{\Gamma_N}\rmd q \,\Dirac[V(q)-Nv],
\end{equation}
or, equivalently, of the entropy
\begin{equation}
s_N(v)=\frac{1}{N}\ln[\Omega_N(v)]
\end{equation}
of a classical system of $N$ degrees of freedom, characterized by a Morse function $V$. The crucial observation for our analysis is that nonanalyticities of $\Omega_N$ are closely related to the {\em critical levels}\/ of $V$, i.\,e., the values $v_\mathrm{c}$ of the potential per degree of freedom $V/N$ for which a critical point $q_\mathrm{c}$ exists such that $V(q_\mathrm{c})/N=v_\mathrm{c}$. If $[v_1,v_2]$ is an interval free of critical levels, one can show that $\Omega_N$ and $s_N$ are smooth functions in this interval (Lemma 2 of \cite{FraPeSpi:07}).

The Morse property \eref{eq:Morse} guarantees that all critical points of $V$ are isolated. We thus may reduce the discussion to the effect of a {\em single}\/ critical point $q_\mathrm{c}$ of $V$ on the analyticity properties of the density of states $\Omega_N$. Without loss of generality we can choose $V(q_\mathrm{c}) = 0$. Furthermore, making use of the {\em Morse lemma} \cite{Demazure,Matsumoto}, a local coordinate system $x=(x_1, x_2, \dots , x_N)$ can be chosen such that
\begin{equation}\label{morse_chart_general}
V[q(x)] = -\sum_{i=1}^{k} x_i^2 + \sum_{i=k+1}^{N} x_i^2
\end{equation}
in some open neighbourhood of $q_\mathrm{c}=q(0)$. We denote the Jacobian of the transformation into this coordinate system by ${\mathfrak J(x)}$ and its determinant by $J(x)$. We will use the expansion of $J(x)$ at $x=0$ for the derivation of the singular behaviour of $\Omega_N(v)$ at $v=0$,
\begin{equation}\label{01}
J(x)=\sum_{I=\{i_1,...,i_N\}}a_I x^I,
\end{equation}
using a multi-index notation $x^I=x_1^{i_1}x_2^{i_2}\cdots x_N^{i_N}$. The zeroth order $I=0$ is particularly symmetric and leads to the leading nonanalytic behaviour of $\Omega_N$. In this section we will restrict the discussion to $I=0$. The analysis of higher order corrections demands different methods and is shifted to the Appendix. The coordinate transformation at $q_\mathrm{c}=0$ is linked to the second derivatives of the potential. Thus $J(0)$ can be written as
\begin{equation}\label{j_hessian}
J(0) = a_0=\left| \det[{\mathfrak H}_V(0)/2] \right|^{-1/2}.
\end{equation}
The determinant $\det{\mathfrak H}_V$ can be viewed as a measure of some ``curvature'' at a critical point, and this allows us to interpret $J$ in our later results as quantifying the flatness of a critical point.

The index $k$ in Eq.\ \eref{morse_chart_general} is called the {\em Morse index} of the critical point. It equals the number of negative eigenvalues of the Hessian ${\mathfrak H}_V(0)$ of $V$ at the critical point $q_\mathrm{c}=0$ and it determines whether a critical point is a minimum ($k=0$), a maximum ($k=N$), or a proper saddle ($k\in\{1,\dots,N-1\}$). In this section we will mainly deal with the latter case. The simpler calculation for the extrema is entailed in the more general set-up of the Appendix. Following the notation in \cite{FraPe:07}, we introduce new variables $X>0$ and $Y>0$ by defining
\begin{equation}\label{0}
X^2 = \sum_{i=1}^{k} x_i^2,\qquad Y^2 = \sum_{i=k+1}^{N} x_i^2,
\end{equation}
yielding
\begin{equation}\label{morse_chart_general_2}
V = - X^2 + Y^2.
\end{equation}
We restrict the calculation of $\Omega_N$ to a ball of radius $r$ centered at $x=0$, 
\begin{equation}\label{1}
X^2 + Y^2\leqslant r^2,
\end{equation}
and we choose $r$ small enough such that the ball fits into the open neighbourhood of Morse's lemma.\footnote{The precise shape of the region is not essential for the calculation.} The restriction to the ball will result in spurious singularities at $Nv=\pm r^2$ which are the limiting values of $V$ on the ball. However, we know that the relevant nonanalyticity has to be located at $v=0$ and thus may assume $N|v|< r^2$ in the following.

Introducing polar coordinates for $(x_1,...,x_k)$ and $(x_{k+1},...,x_N)$, we find for the density of states in the vicinity of $q_\mathrm{c}$ to leading order the expression
\begin{eqnarray}
\Omega_N^{(q_\mathrm{c})}(V) &= J(0) \int {\mathrm d}^N x\,\Dirac \left( -X^2 + Y^2 - V \right) \Heaviside\left( r^2 - X^2-Y^2 \right)\label{3}\\
&= J(0) C_{k-1} C_{N-k-1} \int_0^{\infty} {\mathrm d}X \int_0^{\infty} {\mathrm d}Y X^{k-1} Y^{N-k-1}\nonumber\\
&\quad\times\Dirac \left( -X^2 + Y^2 - V \right) \Heaviside\left( r^2 - X^2-Y^2 \right),\label{3b}
\end{eqnarray}
where $C_n$ is the volume of the $n$-sphere,
\begin{equation}\label{c_n}
C_n = \frac{2 \pi^{(n+1)/2}}{\Gamma \left[(n+1)/2 \right]},
\end{equation}
and
\begin{equation}
\Heaviside(x)=\cases{0 & \mbox{if $x<0$,}\\1 & \mbox{if $x\geqslant0$,}}
\end{equation}
denotes the Heaviside step function. We evaluate the $X$-integral making use of
\begin{equation}
\Dirac(-X^2 + Y^2 - V )=\Dirac[2(Y^2 - V)^{1/2}(X-[Y^2 - V]^{1/2})]\Heaviside(Y^2-V),
\end{equation}
obtaining
\begin{equation}
\Omega_N^{(q_\mathrm{c})}(V)= J(0) C_{k-1} C_{N-k-1} \cases{I_- & \mbox{for $V<0$},\\ I_+ & \mbox{for $V\geqslant 0$},}
\end{equation}
with
\begin{equation}\label{omega_n_beginning}
I_\pm=\frac{1}{2} \int_{\sqrt{V} \; \Heaviside(V)}^{\sqrt{(r^2+V)/2}} {\mathrm d}Y \left( Y^2 - V \right)^{(k-2)/2} Y^{N-k-1}.
\end{equation}
In order to avoid cuts while comparing the two integrals $I_-$ and $I_+$, we consider $V$ as a complex variable and shift it slightly, say, above the real axis,
\begin{equation}\label{4}
\Ima V=\epsilon>0.
\end{equation}
We use the variable transformation
\begin{equation}\label{5}
Y=\sqrt{Vy}
\end{equation}
where we interpret the square root as a single valued function on the complex plane cut from $-\infty$ to 0 (and $\sqrt{1}=+1$). With this definition of the square root we resolve the ambiguities in raising complex numbers to half integer powers. The integrals $I_\pm$ are transformed into
\numparts
\begin{eqnarray}
I_-=\frac{V}{4} \int_{0}^{(r^2+V)/(2V)}\rmd y \sqrt{V(y-1)}^{k-2} \sqrt{Vy}^{N-k-2},\label{6a}\\
I_+=\frac{V}{4} \int_{1}^{(r^2+V)/(2V)}\rmd y \sqrt{V(y-1)}^{k-2} \sqrt{Vy}^{N-k-2}.\label{6b}
\end{eqnarray}
\endnumparts
Since $\Ima V>0$, the upper limit of the integral lies in the lower half plane $H_-$. The integrand is holomorphic in $H_-$, allowing us to shift the contour freely in the lower half plane. The contour is supposed to only touch the real line in the lower limit $y=1$ or $y=0$, respectively. By our convention for the square root we have $\sqrt{y-1}=-\rmi \sqrt{1-y}$ in $H_-$. If, in the case of $I_+$, we let the contour take a detour via $0-\rmi \epsilon$, we are able to express both integrals in terms of incomplete beta functions B$(k_1,k_2,z)$,
\numparts
\begin{eqnarray}
I_+&=&\frac{\sqrt{V}^{N-2}}{4\rmi^{k-2}} \left[{\mathrm B}\left(\frac{N-k}{2},\frac{k}{2},\frac{Y^2+V}{2V}\right)-{\mathrm B}\left(\frac{N-k}{2},\frac{k}{2},1\right)\right],\label{7a}\\
I_-&=&\frac{\sqrt{V}^{N-2}}{4\rmi^{k-2}} {\mathrm B}\left(\frac{N-k}{2},\frac{k}{2},\frac{Y^2+V}{2V}\right).\label{7b}
\end{eqnarray}
\endnumparts
It is also possible to express the integrals in terms of beta functions with arguments in the interval $(0,1)$. To achieve this for $I_-$, we have to apply a M\"obius transform $y=z/(z-1)$ which permutes the points $(0,1,\infty)$ to $(0,\infty,1)$. For the transformation of $I_+$ we have to permute $(0,1,\infty)$ to $(\infty,0,1)$, facilitated by $y=1/(1-z)$. Performing these substitutions we obtain
\numparts
\begin{eqnarray}
I_+&=&\frac{V^{(N-2)/2}}{4}{\mathrm B}\left(\frac{k}{2},1-\frac{N}{2},\frac{r^2-V}{r^2+V}\right),\label{moebius1}\\
I_-&=&\frac{(-V)^{(N-2)/2}}{4}{\mathrm B}\left(\frac{N-k}{2},1-\frac{N}{2},\frac{r^2+V}{r^2-V}\right).\label{moebius2}
\end{eqnarray}
\endnumparts
We obtain for the difference of $I_+$ and $I_-$
\begin{equation}\label{8}
I_+-I_-=\frac{\sqrt{V}^{N-2}}{4\rmi^k} \frac{\Gamma(k/2)\Gamma((N-k)/2)}{\Gamma(N/2)},
\end{equation}
where the complete beta function has been expressed in terms of the $\Gamma$-function.

What are the implications of this result for the nonanalyticity of $I(V)=\Heaviside(V)I_++\Heaviside(-V)I_-$? We know that $I(V)$ is a real function which may have a square root singularity (for odd $N$) or a logarithmic singularity (for even $N$) at $V=0$ [this follows for example from the properties of the beta functions in Eqs. \eref{moebius1} and \eref{moebius2}].
 
In the case of even $k$, the difference $I_+-I_-$ is real for $V>0$ and we simply conclude that $I(V)$ is analytic up to a term $(-1)^{k/2}V^{(N-2)/2}\Heaviside(V)\Gamma(k/2)\Gamma((N-k)/2)/[4\Gamma(N/2)]$.

In the case of odd $k$ and odd $N$, we find that $I_+-I_-$ is real for $V<0$: We have to keep in mind that for $\Ima V>0$ we have $\sqrt{-V}=-\rmi\sqrt{V}$ to see that $I(V)$ is analytic up to a term $(-1)^{(N-k)/2}(-V)^{(N-2)/2}\Heaviside(-V)\Gamma(k/2)\Gamma((N-k)/2)/[4\Gamma(N/2)]$.
 
Finally, in the case $k$ odd and $N$ even, $I_+-I_-$ is imaginary for positive and negative $V$. This is a consequence of a logarithmic singularity which has to be present in both, $I_+$ and $I_-$ (otherwise Eq.\ \eref{8} would contain an $\ln(V)$-term). In fact, for $\Ima V>0$ a term $\ln|V|$ in $I(V)$ results in a term $\ln|V|+\rmi\arg(V)$ in $I_+$ and a term $\ln|V|-\rmi[\pi-\arg(V)]$ in $I_-$ which gives $\rmi\pi$ in $I_+-I_-$. Comparison with Eq.\ (\ref{8}) yields that $I(V)$ is analytic up to a term $(-1)^{(k+1)/2}V^{(N-2)/2}(\ln|V|)\Gamma(k/2)\Gamma((N-k)/2)/[4\pi\Gamma(N/2)]$. We confirm that the nonanalyticity is independent of the radius $r$ of the domain under consideration.

The leading order nonanalyticity depends on the remainder of the index $k\,\mod\,4$ only. This remarkable fact prevails to all orders as we will show in the Appendix. Moreover, the type of the nonanalyticity depends only on whether $N$ and $k$ are even or odd.

Including the prefactors in Eq.\ (\ref{3b}) and going back to $v=V/N$, we obtain the leading order behaviour of the density of states $\Omega_N$ as summarized in the following proposition.
\begin{proposition}\label{prop:finite}
Let $V:G\to\RR$ be a Morse function with a single critical point $q_\mathrm{c}$ of index $k$ in an open region $G$. Without loss of generality, we assume $V(q_\mathrm{c})=0$. Then the density of states $\Omega_N(v)$ is nonanalytic at $v=0$. The leading order nonanalyticity is given by a function depending on the number of degrees of freedom $N$, the index of the critical point $k\,\mod\,4$, and the potential energy per particle $v$. We have
\begin{equation}\label{eq:Omega_sep}
\Omega_N^{\rm na}(v)=\frac{(N\pi)^{N/2}}{N\Gamma(N/2)\sqrt{\left|\det\left[{\mathfrak H}_{V}(q_\mathrm{c})/2\right]\right|}}h_{N,k\,{\rm mod}\,4}^{\rm na}(v)
\end{equation}
with the universal function
\begin{equation}\label{eq:h_Nk}
\fl h_{N,k\,{\rm mod}\,4}^{\rm na}(v)=\cases{
(-1)^{k/2} \,v^{(N-2)/2} \Heaviside(v) & \mbox{for $k$ even,}\\
(-1)^{(k+1)/2} \,v^{(N-2)/2}\,\pi^{-1}\ln|v| & \mbox{for $N$ even, $k$ odd,}\\
(-1)^{(N-k)/2} (-v)^{(N-2)/2} \Heaviside(-v) & \mbox{for $N,k$ odd.}\\
}
\end{equation}
\end{proposition}%
In fact, it is possible to give a complete description of the nonanalytic part of the density of states in terms of the Taylor coefficients of the Jacobian determinant \eref{01} at the critical value. We find that only those indices $I= \{i_1,\dots,i_N\}$ contribute to $\Omega_N$ where all $i_k$ ($k=1,\dots,N$) are even. The general situation is summarized in the following theorem.

\begin{theorem}\label{prop:finite1}
Let $V:G\to\RR$ be a Morse function with a single critical point $q_\mathrm{c}$ of index $k$ in an open region $G$ and $V(q_\mathrm{c})=0$. The density of states can be decomposed into an analytic part $\Omega_N^{\rm a}$ and a nonanalytic part $\Omega_N^{\rm na}$,
\begin{equation}
\Omega_N=\Omega_N^{\rm a}+\Omega_N^{\rm na}.
\end{equation}
The nonanalytic part of the density of states is the product of an analytic function $h_{N,k,J}^{\rm a}$ and the universal nonanalytic function $h_{N,k\,{\rm mod}\,4}^{\rm na}$ given in Eq.\ \eref{eq:h_Nk}. The analytic factor depends on the number of degrees of freedom $N$, the index of the critical point $k$, the Jacobian determinant $J$ of the transformation from the original coordinate system to Morse coordinates, and the potential energy per particle $v$. The nonanalytic factor depends on $N$, $k$ mod 4, and $v$,
\begin{equation}
\Omega_N^{\rm na}(v)=h_{N,k,J}^{\rm a}(v)\, h_{N,k\,{\rm mod}\,4}^{\rm na}(v).
\end{equation}
The analytic factor $h_{N,k,J}^{\rm a}(v)$ can be expressed in terms of the expansion coefficients $a_I$ of the Jacobian determinant at the critical value, Eqs.\ (\ref{01}) and (\ref{j_hessian}). With $I_1=i_1+\dots+i_k$, $I_2=i_{k+1}+\dots+i_N$, and $|I|=I_1+I_2$ we have
\begin{equation}\label{02}
\fl h_{N,k,J}^{\rm a}(v)=\sum_{I=\{i_1,\dots,i_N\}}(-1)^{I_\eta}a_{2I}\frac{\pi^{N/2}N^{N/2+|I|-1}\prod_{j=1}^N(2i_j)!}{\Gamma(N/2+|I|)\prod_{j=1}^Ni_j!}\left(\frac{v}{4}\right)^{|I|},
\end{equation}
\begin{equation}
\eta=
\cases{
1&\mbox{if $N$ or $k$ even,}\\
2&\mbox{if $N$ and $k$ odd.}
}
\end{equation}
\end{theorem}
\begin{proof}
Eq.\ (\ref{02}) will be derived in the Appendix. The leading order $I=0$ yields the prefactor in Eq.\ (\ref{eq:Omega_sep}), thus proving Proposition \ref{prop:finite} in the cases $k=0$ and $k=N$. It remains to show that $h_{N,k,J}^\mathrm{a}(v)$ is analytic at $v=0$. To this end, it is sufficient to show that the sum in Eq.\ (\ref{02}) converges in some neighbourhood of $v=0$. We may assume that the Jacobian determinant is analytic in a neighbourhood $U$ of $x=0$, yielding $|a_I|<R^{-|I|}$ where $R$ is the maximum of the $x_i$ in $U$. Using Sterling's formula we find that, in the limit $|I|\to\infty$, the summands in Eq.\ (\ref{02}) are bounded by
\begin{equation}
\left(\frac{\pi}{|I|}\right)^{(N-1)/2}N^{N/2-1}\left(\frac{vN}{R^2}\right)^{|I|},
\end{equation}
yielding $R^2/N$ for the radius of convergence at $v=0$.
\end{proof}

The theorem gives a complete account of the nonanalyticities in $\Omega_N$ as the consequence of a {\em single}\/ critical point of a Morse function $V$. In the presence of several isolated critical points, their nonanalytic contributions simply have to be added up. It is also possible to state the result in terms of the potential $V$ in the original variables, not making use of Morse coordinates. The result, however, is more complicated and for our purposes the above version will be sufficient.

One can verify that in any of the three cases in \eref{eq:h_Nk}, $\Omega_N$ is $\left\lfloor(N-3)/2\right\rfloor$-times continuously differentiable. This result is in agreement with the nonanalytic behaviour of the exact solution for the density of states of the mean-field spherical model reported in \cite{KaSchne:06}. In other words, the density of states $\Omega_N$ becomes ``smoother'' with increasing number of degrees of freedom, and already for moderate $N$ it will supposedly be impossible to observe such finite-system nonanalyticities from noisy experimental or numerical data. At first sight one might therefore suspect that the nonanalyticities of the entropy are irrelevant for large systems and have no effect in the thermodynamic limit, but the following considerations will show that such an assertion is premature.

\section{Thermodynamic limit for the density of states}
\label{sec:tdl}

In the last years, quite a few articles have been published on the relation between nonanalyticities of the entropy or the free energy in the thermodynamic limit and topology changes in configuration space (see \cite{Kastner:08,Pettini} for a review). The object of study in this approach is the family $\{{\mathcal M}_v\}_{v\in\RR}$ with
\begin{equation}\label{eq:Mv}
{\mathcal M}_v=\left\{q\in\Gamma_N\,\big|\,V(q)\leqslant Nv\right\},
\end{equation}
i.\,e., the subsets of all points $q$ from configuration space $\Gamma_N$ for which the potential energy per particle, $V(q)/N$, is equal to or smaller than a given value $v$. For several models, topology changes of ${\mathcal M}_v$ were studied under variation of the parameter $v$, finding that in many cases the occurrence of a phase transition is signalled by a signature in the Euler characteristic of ${\mathcal M}_v$ (which is a topological invariant) \cite{CaPeCo:03,Angelani_etal:03}. Then, in a recent letter \cite{FraPe:04}, Franzosi and Pettini proved that, for a certain class of short-range models, a topology change in $\{{\mathcal M}_v\}_{v\in\RR}$ at $v=v_\mathrm{t}$ is {\em necessary}\/ for a phase transition to take place at a transition potential energy $v_\mathrm{t}$. This result demonstrates that a relation between nonanalyticities of the entropy or the free energy in the thermodynamic limit and topology changes in configuration space does exist at least for this class of short-range models.

It is a central proposition of Morse theory \cite{Hirsch,Matsumoto} that, for potentials $V$ having the Morse property, each topology change of ${\mathcal M}_v$ at some value $v=v_\mathrm{c}$ corresponds to one or several critical points $q_\mathrm{c}$ of $V$ with critical value $v_\mathrm{c}=V(q_\mathrm{c})/N$. More precisely, if $V$ is a Morse function and if we know all the critical points of $V$ and their critical indices, the handle decomposition theorem \cite{Matsumoto} asserts that we have all the information readily available to specify the topology changes that occur in the family $\{{\mathcal M}_v\}_{v\in\RR}$ of configuration space subsets. As a consequence, the study of topology changes within $\{{\mathcal M}_v\}_{v\in\RR}$ and the study of saddle points of $V$ are closely related approaches. From the proven relation between phase transitions and topology changes \cite{FraPe:04} we can therefore conclude that a relation between saddle points of $V$ and nonanalyticities of the entropy exists not only for finite systems (as worked out in Sec.~\ref{sec:finite}), but also in the thermodynamic limit of infinite system size.

In light of our finite-system results summarized in Proposition \ref{prop:finite} and Theorem \ref{prop:finite1}, this is a remarkable and somewhat surprising observation: Despite their decreasing strength for large system sizes $N$, nonanalyticities of the finite-system entropy appear to be related to their infinite-system counterparts in some way. This relation, however, is not one-to-one: critical points of $V$ are necessary, but by no means sufficient for a phase transition to occur. For several models studied, the number of critical levels $v_\mathrm{c}$ of the potential $V$ was found to increase unboundedly with the number $N$ of degrees of freedom of the system, and the levels become dense on some interval in the thermodynamic limit \cite{CaPeCo:03,Angelani_etal:03,RiRiSta:05}. Therefore, most of the critical points of $V$ are {\em not}\/ related to the phase transition. From these observations, the question arises how, and under which conditions, nonanalyticities of the finite-system entropy may give rise to a phase transition in the thermodynamic limit. This issue will be addressed in the present section.%\footnote{An attempt to address this question has been made in \cite{FraPe:07}, focusing on the contribution to $\Omega_N$ which can{\em not}\/ be at the origin of a phase transition. Furthermore, we believe that the results in \cite{FraPe:07} are flawed by conceptual difficulties which will be discussed in detail elsewhere.}

To this purpose, we investigate the density of states $\Omega_N$ in a small interval $(v_0-\epsilon,v_0+\epsilon)$ around some value $v_0$ of the potential energy. In order to quantify the contribution to the entropy caused by the critical points, we split the density of states into two terms,
\begin{equation}\label{eq:splitting}
\Omega_N^{v_0,\epsilon}(v) = A^{v_0,\epsilon}_N(v) + B_N^{v_0,\epsilon}(v).
\end{equation}
$B_N^{v_0,\epsilon}$ contains the nonanalytic contributions, as specified in Theorem \ref{prop:finite1}, from all critical points in the $\epsilon$-neighbourhood of $v_0$,
\begin{equation}\label{eq:B_N}
B_N^{v_0,\epsilon}(v) = \sum_{\{v_\mathrm{c}:|v_\mathrm{c}-v_0|<\epsilon\}}\;\;\sum_{\{q_\mathrm{c}:V(q_\mathrm{c})/N=v_\mathrm{c}\}}\Omega_{N,q_\mathrm{c}}^{\rm na}(v),
\end{equation}
written as a sum over critical values $v_\mathrm{c}$ of $V$ inside the neighbourhood and over critical points $q_\mathrm{c}(v_\mathrm{c})$ corresponding to the respective critical value. Then a smooth function $A^{v_0,\epsilon}_N$ can be chosen such that the $\epsilon$-density of states $\Omega_N^{v_0,\epsilon}$ from \eref{eq:splitting} coincides with the exact density of states $\Omega_N$ inside the interval $(v_0-\epsilon,v_0+\epsilon)$. In the following we will compute a bound on $B_N^{v_0,\epsilon}$ in order to examine under which conditions the nonanalytic terms stemming from the saddle points may yield a non-vanishing contribution to $\Omega_N^{v_0,\epsilon}$, and therefore to $\Omega_N$, in the thermodynamic limit, thereby possibly inducing a phase transition.

In order to perform the thermodynamic limit, we consider the $\epsilon$-entropy per degree of freedom,
\begin{equation}\label{eq:sv}
s^{v_0,\epsilon}(v)=\lim_{N\to\infty}\frac{1}{N}\ln\left[A^{v_0,\epsilon}_N(v)+B_N^{v_0,\epsilon}(v)\right],
\end{equation}
since, at least for a system with short-range interactions, we may expect this quantity to exist \cite{Ruelle}. Within the interval $(v_0-\epsilon,v_0+\epsilon)$, $s^{v_0,\epsilon}$ coincides with the exact entropy
\begin{equation}
s(v)=\lim_{N\to\infty}\frac{1}{N}\ln\Omega_N(v).
\end{equation}
It is important to note that $N^{-1}\ln B_N^{v_0,\epsilon}$ has in general a non-zero thermodynamic limit: For fixed $I$, we find that the summands in Eq.\ (\ref{02}) are exponential in $N$,
\begin{equation}\label{x3}
(-1)^{I_\eta}a_{2I}\frac{(v/4)^{|I|}\prod_{j=1}^N(2i_j)!}{\sqrt{4\pi N}\prod_{j=1}^Ni_j!}(2\pi\rme)^{N/2},
\end{equation}
permitting $B_N^{v_0,\epsilon}$ to give a finite contribution to $s$.\footnote{A similar splitting of the density of states into two terms, one stemming from the vicinities of the critical points, the other one containing the rest, has been used in \cite{FraPe:07}.} Now we would like to deduce the (non)analyticity of the entropy $s$ from the properties of the thermodynamic limit expressions
\numparts
\begin{eqnarray}
a^{v_0,\epsilon}(v)&=\lim_{N\to\infty}\frac{1}{N}\ln A^{v_0,\epsilon}_N(v),\label{eq:a}\\
b^{v_0,\epsilon}(v)&=\lim_{N\to\infty}\frac{1}{N}\ln B_N^{v_0,\epsilon}(v).\label{eq:b}
\end{eqnarray}
\endnumparts
To this end it is instructive to rewrite Eq.\ \eref{eq:sv} as
\begin{equation}\label{eq:sv2}
s^{v_0,\epsilon}(v)=\max\left\{a^{v_0,\epsilon}(v),b^{v_0,\epsilon}(v)\right\}.
\end{equation}
The above equation holds unless $A^{v_0,\epsilon}_N$ and $B_N^{v_0,\epsilon}$ are very closely related (such as $A^{v_0,\epsilon}_N=-B_N^{v_0,\epsilon}$) which is highly non-generic and thus neglected. As a consequence of Eq.\ \eref{eq:sv2}, we cannot expect to draw any conclusions on $s^{v_0,\epsilon}$ from the knowledge of one single function $a^{v_0,\epsilon}$ or $b^{v_0,\epsilon}$, but only from the interplay between both. If, for example, we find a nonanalyticity in $b^{v_0,\epsilon}$, this nonanalytic behaviour may be visible in $s^{v_0,\epsilon}$ in one case, but it may simply be overruled by a larger $a^{v_0,\epsilon}$ in another instance. Furthermore, a nonanalyticity in $s^{v_0,\epsilon}$ may arise from a {\em crossover}\/ between $a^{v_0,\epsilon}$ and $b^{v_0,\epsilon}$ when the dominant part in the maximum changes from $a^{v_0,\epsilon}$ to $b^{v_0,\epsilon}$ (or vice versa).\footnote{This possibility falsifies the reasoning in \cite{FraPe:07} saying that, since a first term from volume splitting is smooth, a nonanalyticity must be due to the second term.}

The ``density splitting'' in \cite{FraPe:07}, similar to our Eq.~\eref{eq:splitting}, raises the hope that one may expect the analytic part in $A^{v_0,\epsilon}_N$ to converge, inside the interval $(v_0-\epsilon,v_0+\epsilon)$, uniformly to an analytic function for well-behaved short-range potentials. Unfortunately we were not able to prove such a result. Furthermore it appears plausible from the discussion in the preceding paragraphs that, at least for short-range models as covered by the theorem in \cite{FraPe:04}, nonanalyticities of the function $b^{v_0,\epsilon}$, stemming from the nonanalytic contributions of the critical points of the potential $V$, are crucial for the occurrence of a phase transition in the thermodynamic limit. As a consequence of this reasoning, we investigate in more detail the thermodynamic limit $N\to\infty$ of $B^{v_0,\epsilon}_N$ in order to find out under which conditions it may give a nonanalytic contribution to the entropy.

When performing this limit, we will for simplicity restrict the discussion to the subsequence with $N=1\pmod 4$ and to the leading order $|I|=0$ (other subsequences and higher orders can be treated in a similar way). Now the functional form of $\Omega^{\rm na}_N$ can be inserted into the expression \eref{eq:B_N}, yielding
\begin{eqnarray}\label{eq:B_N_int}
\fl B_N^{v_0,\epsilon}(v) &= \frac{(N\pi)^{N/2}}{N\Gamma(N/2)} \biggl\{ \sum_{\{v_\mathrm{c}:v_0-\epsilon<v_\mathrm{c}<v_0\}} \!\!\left(v-v_\mathrm{c}\right)^{(N-2)/2}
\biggl[ \sum_{q_\mathrm{c}\in Q_0(v_\mathrm{c})}\!J(q_\mathrm{c}) - \sum_{q_\mathrm{c}\in Q_2(v_\mathrm{c})}\!J(q_\mathrm{c})\biggr]\nonumber\\
\fl &\quad + \sum_{\{v_\mathrm{c}:v_0<v_\mathrm{c}<v_0+\epsilon\}} \left(v_\mathrm{c}-v\right)^{(N-2)/2}
\biggl[ \sum_{q_\mathrm{c}\in Q_1(v_\mathrm{c})}J(q_\mathrm{c}) - \sum_{q_\mathrm{c}\in Q_3(v_\mathrm{c})}J(q_\mathrm{c})\biggr]
\biggr\}\\
\fl &= \frac{(N\pi)^{N/2}}{N\Gamma(N/2)} \biggl\{ \int_{v_0-\epsilon}^{v_0} \rmd v'\left(v-v'\right)^{(N-2)/2} \left[{\mathcal N}_0(v')-{\mathcal N}_2(v')\right]\nonumber\\
\fl &\quad + \int_{v_0}^{v_0+\epsilon} \rmd v' \left(v'-v\right)^{(N-2)/2} \left[{\mathcal N}_1(v')-{\mathcal N}_3(v')\right]\biggr\}.
\end{eqnarray}
Here,
\begin{equation}
Q_\ell(v_\mathrm{c})=\left\{q_\mathrm{c}\,\big|\,V(q_\mathrm{c})/N=v_\mathrm{c} \,\wedge\, k(q_\mathrm{c})=\ell\pmod 4\right\}
\end{equation}
is the set of all critical points on the critical level $v_\mathrm{c}$ which have index $k=\ell \pmod 4$, $k(q_\mathrm{c})$ denotes the index of the critical point $q_\mathrm{c}$, and
\begin{equation}\label{eq:N_j}
{\mathcal N}_\ell(v')=\sum_{q_\mathrm{c}}J(q_\mathrm{c}) \Dirac(V(q_\mathrm{c})/N-v') \Dirac_{\ell,k(q_\mathrm{c})\;\mathrm{mod\,4}}, \quad \ell=0,1,2,3,
\end{equation}
are the distribution functions of the critical points with index $k=\ell\pmod 4$, weighted by their Jacobian determinant $J$. We expect these Jacobian distributions to have a smooth thermodynamic limit in the sense that their integral means
\begin{equation}
\frac{1}{\epsilon}\int_{v}^{v+\epsilon}\rmd v'{\mathcal N}_\ell(v')=\frac{1}{\epsilon}
\sum_{q_\mathrm{c}\in Q_\ell([v,v+\epsilon])}J(q_\mathrm{c})
\end{equation}
have smooth limits relative to the number of critical points in the interval $[v,v+\epsilon]$. This implies that
\begin{equation}\label{x4}
{\mathcal J}_\ell(v)=\frac{\sum\limits_{q_\mathrm{c}\in Q_\ell([v,v+\epsilon])}J(q_\mathrm{c})}
{\sum\limits_{q_\mathrm{c}\in Q_\ell([v,v+\epsilon])}1}
\end{equation}
has a thermodynamic limit of the form
\begin{equation}\label{x2}
{\mathcal J}_\ell(v)=\exp\left[Nj_\ell(v)\right]
\end{equation}
for small $\epsilon>0$. In this case we may express ${\mathcal N}_\ell$ in terms of ${\mathcal J}_\ell$ and the density $c_\ell$ of critical points with
index $\ell$ mod 4,
\begin{equation}
c_\ell(v)=\frac{1}{\epsilon N_\ell}\sum_{q_\mathrm{c}\in Q_\ell([v,v+\epsilon])}1,
\end{equation}
yielding
\begin{equation}
{\mathcal N}_\ell(v)=N_\ell c_\ell(v){\mathcal J}_\ell(v),
\end{equation}
where
\begin{equation}
N_\ell=\sum_{q_\mathrm{c}\in Q_\ell(\RR)}1, \qquad \ell=0,1,2,3,
\end{equation}
is the total number of critical points with index $\ell$ mod 4. Generically this number grows exponentially
\begin{equation}
N_\ell=\exp(N n_\ell),
\end{equation}
for large $N$, although for specific examples (see the spherical model in Sec.~\ref{sec:spherical}) the growth pattern may be different. Collecting all the above information, we obtain an upper bound for the contribution of the critical points of $V$ to the density of states,
\begin{eqnarray}\label{x1}
\fl B_N^{v_0,\epsilon}(v)\leqslant\frac{(N\pi)^{N/2}}{N\Gamma(N/2)}\max\biggl\{
&N_1\int_{v_0-\epsilon}^{v_0} \rmd v'(v-v')^{(N-2)/2}c_1(v'){\mathcal J}_1(v'),\nonumber\\
&N_3\int_{v_0-\epsilon}^{v_0} \rmd v'(v-v')^{(N-2)/2}c_3(v'){\mathcal J}_3(v'),\nonumber\\
&N_2\int_{v_0}^{v_0+\epsilon} \rmd v'(v'-v)^{(N-2)/2}c_2(v'){\mathcal J}_2(v'),\nonumber\\
&N_4\int_{v_0}^{v_0+\epsilon} \rmd v'(v'-v)^{(N-2)/2}c_4(v'){\mathcal J}_4(v')\biggr\}.
\end{eqnarray}
The fact that the ${\mathcal N}_\ell$ enter Eq.\ (\ref{eq:B_N_int}) as a difference may in very smooth setups lead to a $B_N^{v_0,\epsilon}$ that is smaller than the right hand side of Eq.\ \eref{x1} but we will not need this for the arguments presented here.\footnote{The reasoning in \cite{KaSchneSchrei:07} on this issue is erroneous.} Inserting Eq.\ \eref{x2} in the above integrals leads to Laplace integrals of the form
\begin{equation}
f(v')\exp\left(N_\ell\left[\frac{1}{2}\ln|v-v'|+j_\ell(v')\right]\right)
\end{equation}
with some function $f(v')$. In the thermodynamic limit, these integrals can be evaluated by Laplace's method, yielding the maximum of $\frac{1}{2}\ln|v-v'|+j_\ell(v')$ within the domain of integration. Hence, in the thermodynamic limit \eref{eq:b} we obtain
\begin{eqnarray}
\fl b^{v_0,\epsilon}(v) &\leqslant\sqrt{2\pi\rme}+\max\biggl\{\max_{\ell\in \{0,2\},\, v-\epsilon<v'<v}\Bigl[n_\ell+\frac{1}{2}\ln(v-v')+j_\ell(v')\Bigr],\\
\fl&\qquad\max_{\ell\in \{1,3\},\, v<v'<v+\epsilon}\Bigl[n_\ell+\frac{1}{2}\ln(v'-v)+j_\ell(v')\Bigr]\biggr\}\nonumber\\
\fl &\leqslant\frac{1}{2}\ln(\epsilon)+\sqrt{2\pi\rme}+\max_{\ell\in \{0,1,2,3\},\, |v-v'|<\epsilon}\left[n_\ell+j_\ell(v')\right]\label{eq:bound}
\end{eqnarray}
for the contribution of the saddle points in the interval $(v_0-\epsilon,v_0+\epsilon)$.

The implications of this bound on $b^{v_0,\epsilon}$ are the following: From Eq.\ \eref{eq:sv2}, we know that only the larger one of the terms $a^{v_0,\epsilon}$ and $b^{v_0,\epsilon}$ contributes to the entropy within the interval $(v_0-\epsilon,v_0+\epsilon)$. Now, if the $\max$-term in \eref{eq:bound} is finite, we can always choose $\epsilon$ sufficiently small such that the bound \eref{eq:bound} for $b^{v_0,\epsilon}$ is smaller than $a^{v_0,\epsilon}$. Therefore, the entropy $s$ in the interval $(v_0-\epsilon,v_0+\epsilon)$ is exclusively determined by $a^{v_0,\epsilon}$, which we assume to be a smooth function. This argument can be repeated for any value of $v_0$, thereby demonstrating that the saddle point contributions in the $b$-term are harmless in the sense that they do not lead to nonanalyticities in the entropy. This reasoning, however, relies on the assumption that the $\max$-term in \eref{eq:bound} is finite, and it breaks down in case $n_\ell$ or $j_\ell$ diverge for any $\ell\in \{0,1,2,3\}$. We summarize these observations in the following theorem.

\begin{theorem}\label{thm2}
The saddle point contribution $b^{v_0,\epsilon}(v)$ cannot induce a phase transition at any potential energy in the interval $(v_0-\epsilon,v_0+\epsilon)$ if
\begin{enumerate}
\item the number of critical points is bounded by\/ $\exp(CN)$ for some\/ $C>0$ and\label{cond1}
\item the Jacobian densities (\ref{x4}) have a thermodynamic limit of the form (\ref{x2}) with\/ $j_\ell<\infty$ $\forall \ell\in\{0,1,2,3\}$\label{cond2}
\end{enumerate}
inside the given interval.
\end{theorem}
\begin{proof}
For the leading order, the proof was already sketched above. The leading order describes the 'sizes' of the saddle points, whereas higher orders account for their detailed shapes. By the geometrical meaning of the term $B_N^{v_0,\epsilon}$ it seems implausible for a singularity to be caused by the shape and not by the size of the saddle points. We do not attempt to give a full mathematical proof for this intuition (abusing the term 'theorem'). The result will be corroborated by the examples in Secs.\ \ref{sec:XY} and \ref{sec:k-trig} where it is shown that a phase transition is caused by a divergent Jacobian density $j_\ell(v)$ at the potential energy of the phase transition $v=v_\mathrm{t}$.
\end{proof}
Geometrically, a divergent Jacobian density signals that the corresponding saddle points become in some sense ``asymptotically flat'' in the thermodynamic limit. To illustrate the implications of these considerations, we calculate in the following sections the Jacobian densities $j_\ell$ for various spin models in order to check whether the requirements of Theorem \ref{thm2} are met. For the spherical model with nearest-neighbour interactions, our results will help us to explain the puzzling observations made by Risau-Gusman {\em et al.}\ \cite{RiRiSta:05} that, for this model, the topological signatures observed do not coincide with the phase transition energy. For the mean-field $XY$ model and the mean-field $k$-trigonometric model, we will show that a phase transition occurs precisely at the value of $v$ for which condition (\ref{cond2}) of Theorem \ref{thm2} breaks down.

\section{Spherical model with nearest-neighbour coupling}
\label{sec:spherical}

This model was introduced by Berlin and Kac \cite{BerKac:52} as an exactly solvable caricature of the Ising model of a ferromagnet. Its configuration space $\Gamma_N$ is an $N$-sphere with radius $\sqrt{N+1}$,\footnote{We shift $N$ to an unconventional $N+1$ to match notation with the previous sections.} and the potential is
\begin{equation}\label{eq:V_sph}
V_{\rm sph}:\Gamma_N\to\RR,\qquad q\mapsto -\frac{1}{2}\sum_{i,j=0}^N K_{ij}q_i q_j.
\end{equation}
The degrees of freedom $q_i$ of the model are placed on a $d$-dimensional cubic lattice with $N+1=L^d$ lattice sites and periodic boundary conditions. We consider a coupling matrix $K$ with elements $K_{ij}=1$ when $i$ and $j$ are nearest-neighbouring sites on the lattice, and $K_{ij}=0$ otherwise. The spherical model with nearest-neighbour coupling is solvable in the thermodynamic limit for arbitrary $d$, and a phase transition from a ferromagnetic phase at low temperatures to a paramagnetic phase at high temperatures (or energies) occurs for all $d\geqslant3$ \cite{Joyce}. Since the coupling matrix $K$ is symmetric, it can be diagonalized by a constant orthogonal transformation,
and we can rewrite the potential in the form
\begin{equation}\label{eq:V_sph_diag}
V_{\rm sph}:\Gamma_N\to\RR,\qquad x\mapsto - \frac{1}{2} \sum_{i=0}^N \lambda_i x_i^2,
\end{equation}
where $\lambda_i$ ($i=0,\dots,N$) are the eigenvalues of $K$. As discussed in \cite{RiRiSta:06}, these eigenvalues can be written as
\begin{equation}
\lambda(p_1,\dots,p_d)=2\sum_{\ell=1}^d \cos\left(\frac{2\pi p_\ell}{L}\right),\qquad p_\ell=0,\dots,L-1,
\end{equation}
and we consider the $\lambda_i$ in \eref{eq:V_sph_diag} as the values of $\lambda(p_1,\dots,p_d)$, ordered from the largest to the smallest. Several of these eigenvalues are degenerate, and we define the spectral density of $K$ (as a function of the potential energy per particle $v$) as
\begin{equation}\label{c(v)_finite}
\sigma_N(v) = \frac{1}{N+1} \sum_{p_1,\dots,p_d=0}^{L-1} \delta(\lambda(p_1,\dots,p_d)-v).
\end{equation}

In two recent publications, Risau-Gusman {\em et al.}\ \cite{RiRiSta:05,RiRiSta:06} reported a study of the topology of the subsets ${\mathcal M}_v\subseteq\Gamma_N$ [as defined in \eref{eq:Mv}] of the spherical model with nearest-neighbour coupling. Computing the deformation retract of the ${\mathcal M}_v$, they found that their topology can be characterized by means of the spectral density \eref{c(v)_finite}. Therefore a signature of the topology changes in the limit of large system size should be visible in the thermodynamic limit $N+1=L^d\to\infty$ of the integral mean of the spectral density $\sigma_N$. It was shown in \cite{RiRiSta:06} that this mean converges to
\begin{equation}\label{eq:cv1}
\lim_{N\to\infty}\sigma_N(v)=\frac{1}{\pi}\int_0^\infty \rmd x \cos(2xv)[J_0(x)]^d,
\end{equation}
where $J_0$ denotes the Bessel function of order zero. Depending on the spatial dimension $d$, this limit function has one or several nonanalytic points. However, none of these coincides with the phase transition energy $v_\mathrm{t}$ (see Figure 1 of Reference \cite{RiRiSta:06} for a plot of these functions for various values of $d$). Therefore no signature of the phase transition of the spherical model is seen in the topological quantities studied. 

Inspired by the analysis in \cite{RiRiSta:05,RiRiSta:06}, we study the subsets ${\mathcal M}_v$ as defined in \eref{eq:Mv} by means of Morse theory. Although the potential $V_{\rm sph}$ of the spherical model in \eref{eq:V_sph_diag} is not a Morse function, it can be transformed into one by adding---in the spirit of Eq.~\eref{eq:perturbation}---perturbations $h_i x_i^2$ with small $h_i\in\RR$,
\begin{equation}\label{eq:V_sph_pert}
\bar{V}_{\rm sph}:\Gamma_N\to\RR,\qquad q\mapsto - \frac{1}{2} \sum_{i=0}^N \bar{\lambda}_i x_i^2,
\end{equation}
where $\bar{\lambda}_i=\lambda_i+h_i$ ($i=0,\dots,N$). For convenience, the $h_i$ are chosen such that $\bar{\lambda}_i>\bar{\lambda}_j$ for $i<j$. For this modified potential we want to calculate the Jacobian densities ${\mathcal J}_\ell$ as given in \eref{x4} when performing simultaneously the zero-field limit $h\to0$ and the thermodynamic limit $N\to\infty$. To this end, we consider $h_i$ as $N$-dependent quantities, vanishing in the thermodynamic limit in a suitable way.

As a first step, we determine the critical points and critical indices of $\bar{V}_{\rm sph}$, using the method of Lagrange multipliers. In the Lagrange function
\begin{equation}
{\mathcal L}=- \frac{1}{2} \sum_{i=0}^N \bar{\lambda}_i x_i^2 + \mu\Bigl[\sum_{i=0}^N x_i^2 -(N+1)\Bigr]
\end{equation}
the second term on the right hand side takes into account the spherical shape of the configuration space of the model, and $\mu$ is the corresponding Lagrange multiplier. The critical points of $\bar{V}_{\rm sph}$ are obtained as the solutions of
\begin{equation}
\frac{\partial {\mathcal L}}{\partial x_i}=-\bar{\lambda}_ix_i+2\mu x_i=0\qquad\forall i=0,\dots,N.
\end{equation}
Simple algebra shows that $2(N+1)$ critical points $x_\mathrm{c}$ exist which can be written as
\begin{equation}
x_\mathrm{c}^{(\pm k)}=\pm\sqrt{N+1}\,e_k,\qquad k=0,\dots,N,
\end{equation} 
where $e_k$ is the unit vector in $k$-direction. Inserting this expression into the potential \eref{eq:V_sph_pert}, we obtain the corresponding critical values
\begin{equation}\label{eq:v_lambda}
v_k=\frac{\bar{V}_{\rm sph}(x_\mathrm{c}^{(\pm k)})}{N}=-\frac{\bar{\lambda}_k}{2}(1+N^{-1}),\qquad k=0,\dots,N,
\end{equation}
and we note that precisely two critical points are located at each critical value. The density of critical points is thus linked to the spectral density $\sigma_N$, and in the limit $N+1=L^d\to\infty$ the integral mean of the density of critical points is
\begin{equation}\label{eq:cv}
c(v)=\lim_{N\to\infty}\sigma_N\left(-\frac{v}{2}\right)=\frac{1}{\pi}\int_0^\infty \rmd x \cos(xv)[J_0(x)]^d,
\end{equation}
similar to Eq.\ \eref{eq:cv1}.
 
Now we fix a certain value of $k\in\{0,\dots,N\}$. A good set of coordinates in the vicinity of the critical values $x_\mathrm{c}^{(\pm k)}$ is given by the $x_i$ with $i\neq k$. Solving the spherical constraint
\begin{equation}
\sum_{i=0}^N x_i^2 -(N+1)=0
\end{equation}
for $x_k$ and inserting the resulting expression into the potential \eref{eq:V_sph_pert}, we obtain
\begin{equation}
\bar{V}_{\rm sph}(x)=- \frac{1}{2} \sum_{i\neq k} (\bar{\lambda}_i-\bar{\lambda}_k) x_i^2- \frac{N+1}{2}\bar{\lambda}_k.
\end{equation}
The potential is quadratic in the $x_i$ and leads to a diagonal Hessian ${\mathfrak H}$ at the critical points with matrix elements
\begin{equation}\label{eq:hessian}
{\mathfrak H}_{ij}(x_\mathrm{c}^{(\pm k)}) = (\bar{\lambda}_k-\bar{\lambda}_i)\Dirac_{ij}\quad\mbox{for}\quad i\neq k.
\end{equation}
Due to the ordering of the $\bar{\lambda}_i$, we can read off from this expression that the index of a critical point $x_\mathrm{c}^{(\pm k)}$ equals $k$. Now we can write the Jacobian determinant in terms of the critical values $v_i$ \eref{eq:v_lambda},
\begin{equation}\label{jacobian}
\fl \bar{J}(x_\mathrm{c}^{(\pm k)}) 
= \bigl| \det[{\mathfrak H}(x_\mathrm{c}^{(\pm k)})/2] \bigr|^{-1/2} 
\!= \prod_{i \neq k} \left|\frac{1+N^{-1}}{v_k - v_i}\right|^{1/2} 
\!\!= \exp\biggl\{\frac{1}{2}\sum_{i \neq k} \ln\left|\frac{1+N^{-1}}{v_k - v_i}\right|\biggr\}.
\end{equation}
Expressing the Jacobian determinant as a function of the potential (per particle) $v$ allows us to formulate the thermodynamic limit in terms of the density $c(v)$ of critical points \eref{eq:cv},
\begin{equation}\label{eq:substitution}
\sum_{\substack{ i=0 \newline i \neq k}}^N \longrightarrow N\int_{-d}^{+d}\rmd v\, c(v),
\end{equation}
where $d$ is the spatial dimension of the lattice on which the spherical model is defined. The range of integration $[-d,+d]$ is determined by the minimum and the maximum values of the potential energy per particle. After some manipulations of the integral obtained from \eref{jacobian} with the substitution \eref{eq:substitution}, we arrive at the Jacobian determinant in the thermodynamic limit as a function of $v$,
\begin{equation}\label{j}
\bar{\mathcal J}(v) = \exp\Bigl\{ -\frac{N}{2\pi} \int_{0}^{\infty} \rmd x \left[ J_0(x) \right]^d f(v,x) \Bigr\}.
\end{equation}
The function $f$ in this expression is given by
\begin{eqnarray}
\fl xf(v,x) = \sin(dx) \ln\left( d^2-v^2 \right) + \sin(xv)\left[ \Ci(x(d+v)) - \Ci(x(d-v)) \right]\nonumber\\ - \cos(xv) \left[ \Si(x(d+v)) + \Si(x(d-v)) \right],
\end{eqnarray}
where $\Ci$ and $\Si$ are the cosine and sine integral functions. $f(v,x)$ is an analytic function for all accessible values $v \in (-d,d)$. As a consequence, also $\bar{\mathcal J}(v)$ is analytic and, in particular, continuous for all $v$.

Observing that for the spherical model both, the index of a critical point $x_\mathrm{c}^{(\pm k)}$ as well as the Jacobian determinant $J(x_\mathrm{c}^{(\pm k)})$, are uniquely determined by the critical values $v_k=\bar{V}_{\rm sph}(x_\mathrm{c}^{(\pm k)})/N$ which decrease monotonously with increasing $j$, we find
\begin{equation}
{\mathcal J}_\ell = \bar{\mathcal J}\qquad\text{and}\qquad c_\ell=c,\qquad\ell=0,1,2,3.
\end{equation}
We thus may apply Theorem \ref{thm2} (with any $C>0$ because $N_\ell\to (N+1)/2$) and conclude that for the spherical model with nearest-neighbour coupling the contributions from the critical points of the potential (or, equivalently, from the topology changes of ${\mathcal M}_v$) to the entropy are negligible in the thermodynamic limit.

At first sight this finding seems to be in conflict with the previous discussion where the relevance of the $b$-term in \eref{eq:sv2} for the occurrence of a phase transition in systems with short-range interactions was emphasized. This apparent contradiction can be understood by noting that the spherical model is not genuinely short-range. Although the interaction term in \eref{eq:V_sph} is restricted to pairs of nearest neighbours, the spherical configuration space renders the interaction effectively long-range. For long-range systems, however, as discussed in \cite{HaKa:05}, the connection between phase transitions and configuration space topology is not valid in general and we cannot assume the function $a^{v_0,\epsilon}$ in \eref{eq:sv2} to give a smooth contribution to the entropy in the thermodynamic limit. Similar to the observations made in \cite{HaKa:05} for a different long-range model, we suspect the entropy $s(v,m)$ as a function of the potential energy $v$ and the magnetization $m$ to be a nonconcave function, such that the nonanalyticity in $s(v)=\max_m s(v,m)$ arises from the maximization over $m$. An analytic calculation of the entropy $s(v,m)$ of the spherical model with nearest-neighbour coupling has been reported in \cite{Behringer:05}, but this analysis omits precisely the region in the $(v,m)$-plane for which a nonconcavity of the entropy might be expected.

Note that, in the presence of long-range interactions, we can not {\em a priori}\/ guarantee a phase transition to be triggered by saddle points of $V$ (or topology changes of ${\mathcal M}_v$), since the conditions of the Franzosi-Pettini-Theorem are not met. Nonetheless, topology changes {\em may}\/ be at the origin of a phase transition even in long-range systems, and we present two such examples in the following sections.

\section{Mean-field $XY$ model}
\label{sec:XY}

The mean-field $XY$ model is a system of $N$ plane rotators described by angular variables $q_i\in[0,2\pi)$ ($i=1,\dots,N$). Each rotator is coupled to each other with equal strength $K>0$, where the interactions are described by the potential
\begin{equation}
\fl V_{XY}:[0,2\pi)^N\to\RR,\qquad q\mapsto \frac{K}{2N}\sum_{i,j=1}^N [1-\cos(q_i-q_j)]-h\sum_{i=1}^N \cos q_i.
\end{equation}
The rotators are subject to an external magnetic field of strength $h\in\RR$ which energetically favours orientations $q_i\approx \pi$ for $h<0$, respectively $q_i\approx 0 \pmod {2\pi}$ for $h>0$. In the limit $N\to\infty$ and $h\to0$, the system has a continuous phase transition.

The critical points of this model and their indices have been analyzed by Casetti {\em et al.}\ \cite{CaPeCo:03} for an arbitrary number $N\in\NN$ of rotators. Critical points of $V_{XY}( q)$ were found to occur for $ q= q^{(\mathrm{c})}\in\{0,\pi\}^N$, i.\,e., for $( q_1,\dots, q_N)$ with all components $ q_i$ ($i=1,\dots,N)$ being either $0$ or $\pi$. Hence, in accordance with condition (\ref{cond1}) of Theorem \ref{thm2}, the number of critical points is growing exponentially with $N$. The elements ${\mathfrak H}_{ij}=\partial^2 V/(\partial  q_i \partial  q_j)$ of the Hessian are also reported in \cite{CaPeCo:03} and, evaluated at a critical point, the diagonal elements of the Hessian ${\mathfrak H}$ can be written as
\begin{equation}
\fl {\mathfrak H}_{ii}( q^{(\mathrm{c})})=\biggl(\frac{K}{N}\sum_{k=1}^N\cos q_k+h\biggr)\cos q_i-\frac{K}{N} = \left[K\left(1-\frac{2n_\pi}{N}\right)+h\right]\cos q_i-\frac{K}{N},
\end{equation}
where $n_\pi$ is the number of $\pi$'s in the sequence $( q_1,\dots, q_N)\in\{0,\pi\}^N$. It has been shown in Appendix A.1 of \cite{CaPeCo:03} that, in the thermodynamic limit $N\to\infty$, the contribution of the off-diagonal elements to the eigenvalues of ${\mathfrak H}$ vanishes. Therefore we can write the Jacobian determinant \eref{j_hessian} at a critical point $ q^{(\mathrm{c})}$ as
\begin{eqnarray}
J( q^{(\mathrm{c})})&=\left| \det[{\mathfrak H}( q^{(\mathrm{c})})/2] \right|^{-1/2}\overset{N\gg1}{\longrightarrow} 2^{N/2}\biggl| \prod_{i=1}^N {\mathfrak H}_{ii}( q^{(\mathrm{c})}) \biggr|^{-1/2}\nonumber\\
&=\left| \frac{K}{2}\left(1-\frac{2n_\pi}{N}\right)+\frac{h}{2} \right|^{-N/2},
\end{eqnarray}
valid for large $N$. Inserting, again from Ref.~\cite{CaPeCo:03}, an expression of $n_\pi$ as a function of the potential energy per degree of freedom $v$,
\begin{equation}
\frac{n_\pi(v)}{N}\overset{N\gg1}{\longrightarrow}\frac{1}{2K}\left(K+h\pm\sqrt{K^2+h^2-2Kv}\right),
\end{equation}
we obtain the Jacobian determinant as a function of $v$, 
\begin{equation}
{\mathcal J}(v)=\left|\frac{4}{K^2+h^2-2Kv}\right|^{N/4},
\end{equation}
valid for large $N$. In the zero-field limit $h\to0$ this expression reduces to
\begin{equation}
{\mathcal J}(v)=\left|\frac{4}{K(K-2v)}\right|^{N/4}.
\end{equation}
Similar to the calculation for the spherical model in Sec.\ \ref{sec:spherical}, the indices of the critical points increase monotonously with their respective critical values. Hence, we can conclude that ${\mathcal J}_\ell={\mathcal J}$ ($\ell=0,1,2,3$) and, making use of Eq.\ \eref{x2}, we obtain
\begin{equation}
j_\ell(v)=\frac{1}{2}\ln 2-\frac{1}{4}\ln[K(K-2v)],\qquad\ell=0,1,2,3.
\end{equation}
\begin{figure}[b]
\psfrag{v}{$v$}
\psfrag{j_l}{$j_\ell$}
\psfrag{0.0}{\small $\!0.0$}
\psfrag{0.1}{\small $\!0.1$}
\psfrag{0.2}{\small $\!0.2$}
\psfrag{0.3}{\small $\!0.3$}
\psfrag{0.4}{\small $\!0.4$}
\psfrag{0.5}{\small $\!0.5$}
\psfrag{1.0}{\small $\!1.0$}
\psfrag{1.5}{\small $\!1.5$}
\psfrag{2.0}{\small $\!2.0$}
\psfrag{2.5}{\small $\!2.5$}
\psfrag{3.0}{\small $\!3.0$}
\includegraphics[width=5cm,angle=270]{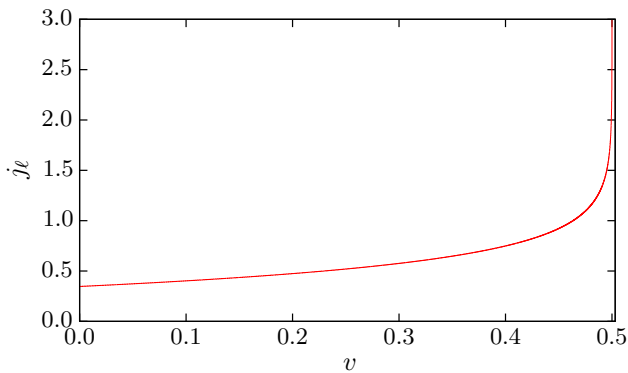}
\caption{\label{fig:jXY}
Plot of the graph of the Jacobian density $j_\ell$ as a function of the potential energy $v$ for the mean-field $XY$ model with coupling constant $K=1$. In agreement with Theorem \ref{thm2}, the transition potential energy $v_\mathrm{t}=1/2$ of this model coincides with the singularity of $j_\ell$.
}
\end{figure}%
The graph of this function is plotted in Fig.~\ref{fig:jXY}. As is easily recognized, $j_\ell(v)$ is singular only at $v=K/2$, which is precisely the value of the potential energy per degree of freedom at which a phase transition occurs for the mean-field $XY$ model with zero external field in the thermodynamic limit. This result nicely illustrates the content of Theorem \ref{thm2}: no phase transition is triggered by the critical points of the potential with a finite Jacobian density. Only those critical points for which $j_\ell$ becomes singular {\em may}\/ induce a phase transition, and this is what apparently happens for the mean-field $XY$ model with $h=0$. Note that, for $h\neq0$ and in the thermodynamic limit, entropic reasons prevent the system from visiting states with potential energy $v=(K^2+h^2)/(2K)$ for which the Jacobian determinant is singular. As a consequence, in accordance with the known thermodynamic behaviour of the mean-field $XY$ model, no phase transition occurs in the presence of a non-zero external field $h$.

Note that, strictly speaking, the analysis of isolated critical points as presented in Secs.~\ref{sec:finite} and \ref{sec:tdl} cannot be applied to the mean-field $XY$ model at the maximum critical level $v=(K^2+h^2)/(2K)$: Precisely at this level the potential $V_{XY}$ is not a proper Morse function, as its critical points are degenerate for arbitrary $N$ and form a critical manifold. However, this should not lead to serious problems for the interpretation of our result: Excluding the trouble-making maximum critical level from our analysis, we can monitor the behaviour of the Jacobian determinant for the second largest critical level. For any finite $N$, the critical points on this level are isolated and hence our analysis is applicable. In the thermodynamic limit, this second largest critical level approaches the problematic maximum one arbitrarily close and the corresponding Jacobian determinant diverges. Note, however, that a divergent Jacobian density need not necessarily be related to the break-down of the Morse property: One might as well imagine a system where all critical points are isolated for all finite $N$, but for which nonetheless the Jacobian density diverges at some point in the thermodynamic limit $N\to\infty$.

\section{Mean-field $k$-trigonometric model}
\label{sec:k-trig}

Another spin model, similar to the one discussed in the previous section, is the mean-field $k$-trigonometric model. A simplified version of this model was introduced in \cite{MaKey:93} as a model of a simple liquid. Like the $XY$ model, the system consists of $N$ plane rotators described by angular variables $q_i\in[0,2\pi)$ ($i=1,\dots,N$). The Hamiltonian function of the mean-field $k$-trigonometric model is
\begin{equation}
\fl V_k:[0,2\pi)^N\to\RR,\qquad  q\mapsto \Delta N^{1-k}\sum_{i_1,\dots,i_k=1}^N [1-\cos(q_{i_1}+\cdots+ q_{i_k})],
\end{equation}
where $q=(q_1,\dots,q_N)$. The constant $\Delta$ determines the coupling strength between the rotators, each interacting with each other at equal strength. Both, thermodynamic behaviour and critical points of the potential have been studied by Angelani {\em et al.}\ \cite{Angelani_etal:05}, and the following results are taken from this reference.

\begin{figure}[hbt]
\psfrag{v}{$v/\Delta$}
\psfrag{j_l}{$j_\ell$}
\psfrag{k=1}{\small $\!\!\!\!\!k=1$}
\psfrag{k=2}{\small $\!\!\!\!\!k=2$}
\psfrag{k=3}{\small $\!\!\!\!\!k=3$}
\psfrag{k=4}{\small $\!\!\!\!\!k=4$}
\psfrag{-0.5}{\small $\!\!\!-0.5$}
\psfrag{ 0.0}{\small $0.0$}
\psfrag{ 0.5}{\small $0.5$}
\psfrag{ 1.0}{\small $1.0$}
\psfrag{ 1.5}{\small $1.5$}
\psfrag{ 2.0}{\small $2.0$}
\includegraphics[width=6cm,angle=270]{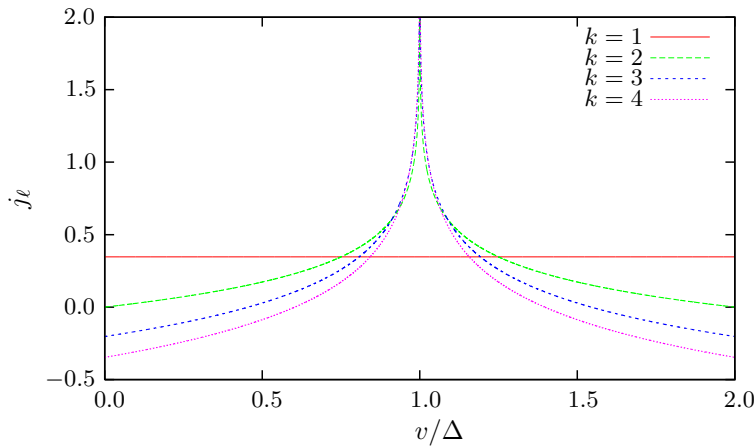}
\caption{\label{fig:jktrig}
Plot of the graph of the Jacobian density $j_\ell$ as a function of the potential energy $v$ (in units of the coupling constant $\Delta$) for the mean-field $k$-trigonometric model with $k=1,2,3,4$. In agreement with Theorem \ref{thm2}, the transition potential energy $v_\mathrm{t}=\Delta$ of this model coincides with the singularity of $j_\ell$ for all values $k\geqslant2$ for which a phase transition takes place.
}
\end{figure}

The parameter $k\in\NN$ in the Hamiltonian function crucially determines the thermodynamic behaviour of the model: In the limit $N\to\infty$, the system shows a discontinuous phase transition for $k\geqslant3$, a continuous one for $k=2$, and no phase transition for $k=1$. Results on the critical points of $V_k$ and on the Hessian at these points can be assembled to obtain the Jacobian density $j_\ell$ as a function of the potential energy per particle $v$,
\begin{equation}
j_\ell(v)=\frac{1}{2}\ln\left(\frac{2}{\Delta k}\left|1-\frac{v}{\Delta}\right|^{1/k-1}\right),\qquad\ell=0,1,2,3.
\end{equation}
The graph of this function is plotted in Fig.~\ref{fig:jktrig} for $k=1,2,3,4$. The calculation is analogous to the one for the mean-field $XY$ model in the previous section and we hence skip the details.

This is another result nicely illustrating the content of Theorem~\ref{thm2}: The function $j_\ell$ is constant (and therefore bounded above) for $k=1$, in agreement with the fact that no phase transition occurs in this case. For $k\geqslant2$, however, the Jacobian density $j_\ell(v)$ shows a divergence at $v=\Delta$, which is precisely the value of the potential energy per degree of freedom at which the mean-field $k$-trigonometric model undergoes a phase transition in the thermodynamic limit.

\section{Summary}
\label{sec:summary}

We have analyzed the relation between saddle points of the potential energy $V$ of classical $N$-particle systems and the analyticity properties of their thermodynamic functions. For finite systems, each saddle point $q_\mathrm{c}$ was found to cause a nonanalyticity in the entropy $s_N(v)$ at the value $v=V(q_\mathrm{c})/N$ of the potential energy, and the functional form of the nonanalytic term is given in Theorem \ref{prop:finite1}. Since the number of saddle points is expected to grow exponentially with $N$ for generic potentials, we arrive at the remarkable conclusion that the finite-system entropy is in general a highly non-smooth function. For large $N$, the order of the nonanalytic term increases unboundedly, leading to an increasing differentiability of $s_N$. Considering the contribution of very many saddle points becoming dense in the thermodynamic limit, we discussed how, despite the ``increasing smoothness'' of $s_N$, a continuous distribution of saddle points with singular Jacobian density may lead to a nonanalyticity in the infinite-system entropy. Interpreting our findings in the spirit of the topological approach to phase transition, our results show under which conditions topology changes of the configuration space subsets ${\mathcal M}_v$ as defined in \eref{eq:Mv} can be at the origin of a phase transition in the thermodynamic limit.

An application of these findings to the spherical model allowed us to understand the puzzling observations of Risau-Gusman {\em et al.}\ \cite{RiRiSta:05,RiRiSta:06} that, for this model, the topological signatures observed do not coincide with the phase transition energy. For the mean-field $XY$ model and the mean-field $k$-trigonometric model we showed that a phase transition occurs precisely at the value of the potential energy per degree of freedom $v$ for which condition (\ref{cond2}) of Theorem \ref{thm2} breaks down. For these two mean-field models, the divergences in the Jacobian densities $j_\ell$ in the thermodynamic limit are foreshadowed by the break-down of the Morse property of the potentials already for finite systems. Note, however, that this is not necessarily the case and one can as well imagine a divergent Jacobian density to emerge in the thermodynamic limit from a potential which is a good Morse function for all finite system sizes $N$.

The criterion for the absence of a phase transition we have presented in Theorem \ref{thm2} is to some extend related to what has become known as the {\em topological hypothesis}\/ in the literature, where a relation between the occurrence of a phase transition and certain ``pronounced'' topology changes of the subsets ${\mathcal M}_v$ [Eq.\ \eref{eq:Mv}] in configuration space was conjectured (see \cite{Kastner:08} for a discussion of the different formulations of this hypothesis). In fact, for the mean-field $XY$ model and the mean-field $k$-trigonometric model in the thermodynamic limit, signatures were found in the Euler characteristic of ${\mathcal M}_v$, occurring precisely at the transition potential energy $v=v_\mathrm{t}$, and a relation between these signatures and the occurrence of phase transitions was conjectured \cite{CaPeCo:03,Angelani_etal:03}. Our Theorem \ref{thm2}, similar to this conjectured relation, has a topological ingredient as well, taking the saddle points of the potential (each of which is related to a topology change of the ${\mathcal M}_v$) as a starting point. In addition to this topological ingredient, we have demonstrated that the local curvatures at the saddle points are important to quantify the effect of these saddle points on thermodynamic quantities. In this way, we {\em derive}\/ a criterion, in part of topological and in part of geometrical nature, which can exclude the occurrence of a phase transition for most of the accessible values of the potential energy $v$, leaving only a few distinct values of $v$ as possible candidates for the occurrence of a phase transition.

Conceptually, our study explores the connection between phase transition theory and the study of energy landscapes, a rapidly developing field with applications, among others, to clusters, biomolecules, and glass-formers \cite{Wales}. One might hope to profit from the considerable knowledge on the relation of energy landscapes and dynamical properties in future work.

\appendix

\section{Proof of Theorem \ref{prop:finite1}}
\label{sec:transformation_omega_n}

In this Appendix we will examine the effect of an $x$-dependent Jacobian $J(x)$ on the singular behaviour of the density of states $\Omega_N$.

Upon Taylor-expanding the Jacobian, we obtain terms proportional to $x^I=x_1^{i_1}x_2^{i_2}\cdots x_N^{i_N}$. These terms result in corrections to Eq.\ (\ref{3}) of the form
\begin{equation}
\Delta_I=a_I\int\rmd^Nx\,x^I\Dirac \left( -X^2+Y^2-V \right) \Heaviside\left( r^2 -X^2-Y^2\right),\label{A1}
\end{equation}
where $a_0=|\det[{\mathfrak H}(q_\mathrm{c})/2]|^{-1/2}$. Due to the symmetry upon substituting $x_j\to -x_j$, we see that only even indices $i_j$ contribute. We use $\Heaviside( r^2 -X^2-Y^2)=\int_0^{r^2}\rmd y\,\Dirac(y-X^2-Y^2)$ and express $\Delta_I$ in terms of a generating function
\begin{eqnarray}
\Delta_I=a_I\int_0^{r^2}\!\!\rmd y\,\rmi^{-|I|}\partial^I_p G(p,y)\Big|_{p=0},\\
\fl G(p,y)=\int\rmd^Nx\,\Dirac \left( -X^2+Y^2-V \right)\Dirac\left(y-X^2-Y^2\right)\exp\Bigl(\rmi\sum_{j=1}^Nx_jp_j\Bigr),\label{A2}
\end{eqnarray}
where we have used the notations $\partial^I_p=\partial^{i_1}_1\partial^{i_2}_2\cdots\partial^{i_N}_N$, $\partial_j=\partial/\partial p_j$, and $|I|=i_1+i_2+...+i_N$. Now we represent the $\Dirac$-distribution by their Fourier components. For convergence we introduce a small $\epsilon>0$ and obtain
\begin{eqnarray}
G(p,y)=\frac{1}{(2\pi)^2}\int\rmd s\int\rmd t\int\rmd^Nx\nonumber\\
\fl \times\exp\Bigl[\rmi s(-X^2+Y^2-V)+(\rmi t+\epsilon)(y-X^2-Y^2)+\rmi\sum_{j=1}^Nx_jp_j-\epsilon(s^2+t^2)\Bigr].\label{A3}
\end{eqnarray}
The argument of the exponential is quadratic in $x$ and can be rewritten as
\begin{eqnarray}
&&\sum_{j=1}^k\left[-(\epsilon+\rmi s+\rmi t)\left(x_j-\frac{\rmi p_j}{2(\epsilon+\rmi s+\rmi t)}\right)^2-\frac{p_j^2}{4(\epsilon+\rmi s+\rmi t)}\right]\nonumber\\
&+&\sum_{j=k+1}^N\left[-(\epsilon-\rmi s+\rmi t)\left(x_j-\frac{\rmi p_j}{2(\epsilon-\rmi s+\rmi t)}\right)^2-\frac{p_j^2}{4(\epsilon-\rmi s+\rmi t)}\right]\nonumber\\
&-&\rmi sV+(\rmi t+\epsilon)y-\epsilon(s^2+t^2).\label{A4}
\end{eqnarray}
From this expression we read off the value of the Gaussian integral and obtain
\begin{eqnarray}
G(p,y)=\frac{\pi^{N/2}}{(2\pi)^2}\int\rmd s\int\rmd t\nonumber\\
\times\frac{\exp\Bigl[-\frac{\sum_{j=1}^kp_j^2}{4(\epsilon+\rmi s+\rmi t)}-\frac{\sum_{j=k+1}^Np_j^2}{4(\epsilon-\rmi s+\rmi t)}
-\rmi sV+(\rmi t+\epsilon)y-\epsilon(s^2+t^2)\Bigr]}{\sqrt{\epsilon+\rmi s+\rmi t}^k\sqrt{\epsilon-\rmi s+\rmi t}^{N-k}},\label{A5}
\end{eqnarray}
using the same definition of the complex square root as in the main text. The generating function factorizes in the variables $p_j$. We use the shorthands $I_1=\sum_{j=1}^ki_j$ and $I_2=\sum_{j=k+1}^Ni_j$ (we have $I_1+I_2=|I|$) and obtain
\begin{eqnarray}
\Delta_I=\frac{a_I\pi^{N/2}\prod_{j=1}^N i_j!}{(2\pi)^2 4^{|I|/2}\prod_{j=1}^N (i_j/2)!}\nonumber\\
\times\int_0^{r^2}\!\!\rmd y\,{\rm e}^{\epsilon y}\int\rmd s\int\rmd t\,
\frac{\exp\left[-\rmi sV+\rmi ty-\epsilon(s^2+t^2)\right]}{\sqrt{\epsilon+\rmi s+\rmi t}^{k+I_1}\sqrt{\epsilon-\rmi s+\rmi t}^{N-k+I_2}}.\label{A6}
\end{eqnarray}
Now we observe that the above double integral in $s$ and $t$ factorizes after changing variables to $u=(t+s)/2$ and $w=(t-s)/2$. We define the function
\begin{equation}\label{A7}
F_j(z)=\frac{1}{2\pi}\int_{-\infty}^\infty \rmd u\,\frac{1}{\sqrt{\epsilon+\rmi u}^j}\,\rme^{\rmi uz-4\epsilon u^2},
\end{equation}
drop the factor ${\rm e}^{\epsilon y}$ which has no effect on the finite integral over $y$ in the limit $\epsilon\to 0$, and replace $\epsilon$ by $2\epsilon$, yielding
\begin{equation}\label{A8}
\Delta_I=\frac{2a_I(\pi/2)^{N/2}\prod_{j=1}^N i_j!}{8^{|I|/2}\prod_{j=1}^N (i_j/2)!}\int_0^{r^2}\!\!\!\rmd y\,F_{k+I_1}(y-V)F_{N-k+I_2}(y+V).
\end{equation}
We have to evaluate $F_0(z)$ for real values of $z$ and non-negative integers $j$. Obviously, $F_0(z)=\Dirac(z)$.

For positive, even $j=2n$ the integrand has a pole at $u=\rmi \epsilon$ of order $n$. If $z<0$ we may evaluate the integral by closing in the lower half plane. We miss the pole and $F_{2n}$ vanishes. If $z>0$ we may evaluate the integral by the residue theorem closing in the upper half plane. Expanding the exponential gives $F_{2n}=\Heaviside(z)z^{n-1}/(n-1)!$.

In the case $j=1$ the integrand has a cut from $u=\rmi \epsilon$ to i${\infty}$. Again we obtain a non-vanishing result only in the case $z>0$ where we may deform the contour to follow the cut. We let $\epsilon$ approach zero and obtain $F_1(z)=\pi^{-1}\Heaviside(z)\int_0^{\rmi \infty}\rmd u\,{\rm e}^{\rmi uz}/\rmi \sqrt{|u|}$. This leads to a Gamma-function evaluated at $1/2$ and we have $F_1=(\pi z)^{-1/2}\Heaviside(z)$.

The case of odd $j=2n+1$ can be reduced to the case $j=1$ by integration by parts. We may drop the factor $\rme^{-4\epsilon u^2}$ in this case and obtain $F_{2n+1}(z)=F_{2n-1}(z)\,2z/(2n-1)$. We iterate this equation and use the above result for $j=1$ to obtain a result analogous to the case of even $j$. This allows us to write the result in compact form
\begin{equation}
F_0(z)=\Dirac(z),\quad F_j(z)=\frac{z^{j/2-1}}{\Gamma(j/2)}\Heaviside(z)\qquad\text{if $j\geqslant 1$}.\label{A9}
\end{equation}
Now we have to distinguish the cases of a maximum, a minimum, or a proper saddle point. In the first case we have $k=I_1=0$ which reduces $F_{k+I_1}(y-V)$ to $\Dirac(y-V)$. We only obtain a non-vanishing result if $0<V<r^2$, and collecting all factors gives
\begin{equation}
\fl \Delta_I=\frac{a_I\pi^{N/2}\prod_{j=1}^N i_j!}{2^{|I|}\Gamma\left[(N+|I|)/2\right]\prod_{j=1}^N(i_j/2)!}V^{(N+|I|-2)/2}\Heaviside(V)\Heaviside(r^2-V)\label{A10}
\end{equation}
for $k=0$. In the case of a minimum we have $k=N$, $I_2=0$ with a non-vanishing result only for $-r^2<V<0$. Upon $V\to-V$ we reproduce---as it should be---the formula of a maximum
\begin{equation}
\fl \Delta_I=\frac{a_I\pi^{N/2}\prod_{j=1}^N i_j!}{2^{|I|}\Gamma\left[(N+|I|)/2\right]\prod_{j=1}^N(i_j/2)!}(-V)^{(N+|I|-2)/2}\Heaviside(-V)\Heaviside(r^2+V)\label{A11}
\end{equation}
for $k=N$. In both cases we confirm the statements of Proposition~\ref{prop:finite} and Theorem~\ref{prop:finite}.

The case of a proper saddle yields
\begin{eqnarray}
\Delta_I&=\frac{2a_I(\pi/2)^{N/2}\prod_{j=1}^N i_j!}{8^{|I|/2}\Gamma\left[(N+I_1)/2\right]\Gamma\left[(N-k+I_2)/2\right]\prod_{j=1}^N (i_j/2)!}\\
&\quad\times\int_{|V|}^{r^2}\!\!\rmd y\,(y-V)^{(k+I_1-2)/2}(y+V)^{(N-k+I_2-2)/2}\label{A12}
\end{eqnarray}
for $0<k<N$. This expression entails the integrals $I_\pm$ from Eq.\ (22) in a symmetrized form. In fact, a transformation $y=V(2z-1)$ allows us to express $\Delta_I$ in terms of $I_\pm$ with $k$ replaced by $k+I_1$ and $N$ replaced by $N+|I|$ (denoted by an upper index),
\begin{equation}
\fl \Delta_I=\frac{4a_I\pi^{N/2}\prod_{j=1}^N i_j!}{2^{|I|}\Gamma\left[(N+I_1)/2\right]\Gamma\left[(N-k+I_2)/2\right]\prod_{j=1}^N (i_j/2)!}I_\pm^{(k+I_1,N+|I|)}.\label{A13}
\end{equation}
Now we can follow the discussion in the main text substituting $(k,N)$ by $(k+I_1,N+|I|)$ when necessary. We observe that, with $I_1$ and $|I|$ being even integers, higher order corrections do not mix the cases in the derivation of Proposition \ref{prop:finite}. Collecting all factors gives the desired result.\\

\bibliographystyle{unsrt}
\bibliography{PTsaddles_long}

\end{document}